
\input phyzzx

\let\refmark=\NPrefmark  
\def\define#1#2\par{\def#1{\Ref#1{#2}\edef#1{\noexpand\refmark{#1}}}}
\def\con#1#2\noc{\let\?=\Ref\let\<=\refmark\let\Ref=\REFS
         \let\refmark=\undefined#1\let\Ref=\REFSCON#2
         \let\Ref=\?\let\refmark=\<\refsend}

\define\RJACKIW
R. Floreanini and R. Jackiw, Phys. Rev. Lett. {\bf 59} (1987) 1873.

\define\RNARAIN
K. Narain, Phys. Lett. {\bf B169} (1986) 41.

\define\RNSW
K. Narain, H. Sarmadi and E. Witten, Nucl. Phys. {\bf B279} (1987) 369.

\define\RGAUNT
J. Gauntlett, J. Harvey and J. Liu, preprint EFI-92-67 (hep-th/9211056).

\define\RMARSCH
N. Marcus and J. Schwarz, Nucl. Phys. {\bf B228} (1983) 145.

\define\RDUALSUGRA
A. Chamseddine, Phys. Rev. {\bf D24} (1981) 3065;
S. Gates and H. Nishino, Phys. Lett. {\bf B173} (1986) 52;
A. Salam and E. Sezgin, Physica Scripta {\bf 32} (1985) 283.

\define\RORT
T. Ortin, preprint SU-ITP-92-24 (hepth@xxx/9208078).

\define\ROLIVE
C. Montonen and D. Olive, Phys. Lett. {\bf B72} (1977) 117;
P. Goddard, J. Nuyts and D. Olive, Nucl. Phys. {\bf B125} (1977) 1.

\define\RROWI
R. Rohm and E. Witten, Ann. Phys. (NY) {\bf 170} (1986) 454.

\define\ROSBORN
H. Osborn, Phys. Lett. {\bf B83} (1979) 321.

\define\RKALORTTWO
R. Kallosh and T. Ortin, preprint SU-ITP-93-3 (hep-th/9302109).

\define\RIBANEZ
A. Font, L. Ib\'a\~nez, D. Lust and F. Quevedo, Phys. Lett. {\bf B249} (1990)
35;
S.J. Rey, Phys. Rev. {\bf D43} (1991) 526.

\define\RHARLIU
J. Harvey and J. Liu, Phys. Lett. {\bf B268} (1991) 40.

\define\RSTROM
A. Strominger, Nucl. Phys. {\bf B343} (1990) 167; C. Callan, J. Harvey and
A. Strominger, Nucl. Phys. {\bf B359} (1991) 611; {\bf B367} (1991) 60;
preprint EFI-91-66 (hep-th/9112030).

\define\RDUFF
M. Duff, Class. Quantum Grav. {\bf 5} (1988) 189;
M. Duff and J. Lu, Nucl. Phys. {\bf B354} (1991) 141;
Phys. Rev. Lett. {\bf 66} (1991) 1402;
Class. Quantum Grav. {\bf 9} (1991) 1;
M. Duff, R. Khuri and J. Lu, Nucl. Phys. {\bf B377} (1992) 281;
J. Dixon, M. Duff and J. Plefka, Phys. Rev. Lett. {\bf 69} (1992) 3009.

\define\RDUFFLUT
M. Duff and J. Lu, Nucl. Phys. {\bf B357} (1991) 354.

\define\RDUFFLU
M. Duff and J. Lu, Nucl. Phys. {\bf B354} (1991) 129.

\define\RGAZU
M. Gaillard and B. Zumino, Nucl. Phys. {\bf B193} (1981) 221.

\define\ROLD
E. Cremmer, J. Scherk and S. Ferrara, Phys. Lett. {\bf B74} (1978) 61;
M. Gaillard and B. Zumino, Nucl. Phys. {\bf B193} (1981) 221;
M. De Roo, Nucl. Phys. {\bf B255} (1985) 515.

\define\RDGHR
A. Dabholkar, G. Gibbons, J. Harvey and F. Ruiz, Nucl. Phys. {\bf B340}
(1990) 33;
A. Dabholkar and J. Harvey, Phys. Rev. Lett. {\bf 63} (1989) 719.

\define\RSTRING
A. Sen, Nucl. Phys. {\bf B388} (1992) 457.

\define\RODD
S. Ferrara, J. Scherk and B. Zumino, Nucl. Phys. {\bf B121} (1977) 393;
E. Cremmer, J. Scherk and S. Ferrara, Phys. Lett. {\bf B68} (1977) 234;
E. Cremmer and J. Scherk, Nucl. Phys. {\bf B127} (1977) 259;
E. Cremmer and B. Julia, Nucl. Phys. {\bf B159} (1979) 141;
E.  Cremmer, J. Scherk and J. Schwarz, Phys. Lett. {\bf B84} (1979) 83.

\define\RMAHSCH
J. Maharana and J. Schwarz, Nucl. Phys. {\bf B390} (1993) 3.

\define\RSCHWARZ
J. Schwarz, preprint CALT-68-1815 (hep-th/9209125).

\define\RTSW
A. Shapere, S. Trivedi and F. Wilczek, Mod. Phys. Lett. {\bf A6}
(1991) 2677.

\define\RDUALITY
A. Sen, preprint TIFR-TH-92-41 (hep-th/9207053) (to appear in Nucl.
Phys. B).

\define\RDYONO
A. Sen, Phys. Lett. {\bf B303} (1993) 22.

\define\RDYONT
A. Sen, preprint NSF-ITP-93-29 (hep-th/9303057).

\define\RTSEYTLIN
A. Tseytlin, Phys. Lett. {\bf B242} (1990) 163; Nucl. Phys. {\bf B350}
(1991) 395.

\define\RHENN
M. Henneaux and C. Teitelboim, p. 79 in Proc. {\it Quantum Mechanics of
Fundamental Systems 2} (Santiago 1987); Phys. Lett. {\bf B206} (1988) 650.

\def\ck{{\cal K}}
\def\tA{\tilde A}
\def\cd{{\cal D}}
\def\hK{\hat K}
\def\ca{{\cal A}}
\def\cb{{\cal B}}
\def\cc{{\cal C}}
\def\hG{\hat G}
\def\hB{\hat B}
\def\cm{{\cal M}}
\def\a{(\alpha)}
\def\cl{{\cal L}}
\def\p{\partial}
\def\aa{(a,\alpha)}
\def\bb{(b,\beta)}

{}~\vbox{\hbox{NSF-ITP-93-46}\hbox{CALT-68-1863}
\hbox{TIFR-TH-93-19}\hbox{hep-th/9304154}\hbox{April
1993}}\break

\title{DUALITY SYMMETRIC ACTIONS}

\author{John H. Schwarz\foot{Supported in part by the U.S. Dept. of
Energy under Grant No. DE-FG03-92-ER40701.}
\foot{JHS@THEORY3.CALTECH.EDU, JHS@NSFITP.ITP.UCSB.EDU}}

\address{Institute for Theoretical Physics, University of California,
Santa Barbara, CA 93106}

\address{California Insitute of Technology,
Pasadena, CA 91125\foot{Permanent address.}}

\andauthor{Ashoke Sen\foot{SEN@TIFRVAX.BITNET,
SEN@SBITP.UCSB.EDU}\foot{Supported
in part by National Science Fundation under grant No. PHY89-04035.}}

\address{Institute for Theoretical Physics, University of California,
Santa Barbara, CA 93106}

\address{Tata Institute of Fundamental Research, Homi Bhabha Road,
Bombay 400005, India${}^{\ddagger}$}

\vfill\eject

\abstract

It is frequently useful to construct dual descriptions of theories
containing antisymmetric
tensor fields by introducing a new potential whose curl gives the dual
field strength, thereby interchanging field equations with Bianchi
identities. We describe a general procedure for constructing actions
containing both potentials at the same time, such that the dual
relationship of the field strengths arises as an equation of motion.
The price for doing this is the sacrifice of manifest Lorentz invariance
or general coordinate invariance, though both symmetries can be realized
nonetheless. There are various examples of global symmetries that have
been realized as symmetries of field equations but not actions. These
can be elevated to symmetries of the action by our method. The main
example that we focus on is the low-energy effective action description
of the heterotic string theory
compactified on a six-torus to four dimensions. We show that the SL(2,R)
symmetry, whose SL(2,Z) subgroup has been conjectured to be an exact
symmetry of the
full string theory, can be realized on the action in a way that brings
out a remarkable similarity to the target space duality symmetry O(6,22).
Our analysis indicates that SL(2,Z)
symmetry may arise naturally in a dual formulation of the theory.

\vfill
\endpage

\chapter{Introduction}

Montonen and Olive\ROLIVE\ conjectured in 1977 that
some theories with a spontaneously broken gauge symmetry possess a
duality symmetry that interchanges electrically charged elementary particles
with magnetically charged t' Hooft--Polyakov monopoles. Such a
symmetry would relate strong coupling to weak coupling, since it sends
the coupling constant to its inverse.
Later analysis showed that among four-dimensional field theories the
best candidate for realizing the Montonen--Olive duality conjecture
is the globally supersymmetric $N=4$ Yang--Mills--Higgs system\ROSBORN.
A similar duality conjecture for ten dimensions
that would relate strong coupling in string
theory to weak coupling in five-brane theory was made in
refs.\con\RSTROM\RDUFF\RDUFFLU\RDUFFLUT\noc.

Apparently unrelated work, at about the
same time as the Montonen--Olive conjecture, showed that many
extended supergravity theories in four dimensions have global
non-compact symmetries\RODD\ROLD.
Some of these symmetries were realized
as symmetries of the
action, whereas others were only demonstrated to be
symmetries of the equations of
motion. In particular, many of these theories
contain an SU(1,1) symmetry (or, equivalently, an SL(2,R) symmetry),
which is a symmetry of the equations of motion only.

Ref.\RMAHSCH\ investigated
dimensional reduction of the bosonic sector of $N=1$ supergravity theory
in ten dimensions, coupled to a set of abelian gauge field
supermultiplets, to four dimensions.
The resulting action describes the bosonic part of the
low-energy effective field theory for the
heterotic string theory compactified on a six-dimensional torus at a
generic point in the moduli space, where all non-abelian symmetries
are broken. The action of this theory has a manifest global
O(6,22) symmetry. A
discrete O(6,22;Z) subgroup of this, which is a symmetry of the Narain
lattice\RNARAIN\RNSW, can be shown to be an exact symmetry
of the compactified
string theory at each order of the string loop perturbation expansion.

This dimensionally reduced theory also turns out to have a hidden
SL(2,R) symmetry,\con\RTSW\RDUALITY\RSCHWARZ\noc\
which is only a symmetry of the
equations of motion and not of the action. Part of this symmetry is
broken by the instanton corrections in string theory. It was conjectured
in refs.\RDUALITY\RSCHWARZ\
that the remaining subgroup, which turns out to be the
discrete group SL(2,Z), may be an exact symmetry of the heterotic string
theory compactified on a six-dimensional torus.
(This suggestion was originally made in the context of a
generic four-dimensional string theory by Ib\'a\~nez et. al.\RIBANEZ\ based
on the analysis of the scalar sector of these theories.)
It was argued in ref.\RDUALITY\ that since
the elementary strings can be regarded as soliton solutions in this
effective field theory\RDGHR\RSTRING, SL(2,Z) invariance of the
effective field theory may be all that is required
to establish SL(2,Z) invariance
of the full string theory.
Further support for SL(2,Z) symmetry in toroidally compactified
heterotic string theory was provided by noting that the spectrum of
electric and magnetic charges, and also the known part of the mass
spectrum of the supersymmetric states in this
theory, are all consistent with the proposed SL(2,Z)
symmetry\RDYONO\RDYONT.

Although SL(2,R)$\times$O(6,22) appears as a symmetry of the classical
equations of motion of the low-energy effective action, the two factors
seem to be on a somewhat different footing:
O(6,22) is a symmetry of the effective action, whereas SL(2,R) is only a
symmetry of the equations of motion. Also, the discrete subgroup
O(6,22;Z) is a symmetry of the string spectrum at the string tree level,
but the discrete subgroup SL(2,Z) is certainly not a symmetry of tree
level string theory (though it could be a symmetry of the full
non-perturbative string theory), since it interchanges string states with
't Hooft--Polyakov-type monopole solutions.

One of the main purposes of this paper is to reformulate the theory in
such a way that, at least in the context of low-energy effective field
theory, the O(6,22) and SL(2,R) symmetries
appear on a more or less equal footing.
In particular, we shall rewrite the dimensionally reduced action in such
a way that both the O(6,22) and the SL(2,R) transformations appear as
symmetries of the action, and not just of the equations of
motion.
The price that must be paid for this is {\it manifest} general coordinate
invariance of the action, though the action does have general coordinate
invariance. The way this works is that the action is invariant under a
symmetry that reduces to the usual general coordinate transformations when
certain auxiliary fields are eliminated by their equations of motion.
Actually, spatial reparametrization invariance remains manifest.

We start in sect.2 with a very simple system that illustrates the key
new feature of our construction, namely, free Maxwell
theory. In the usual formulation,
the equations of motion of Maxwell theory (without sources)
are symmetric under the
duality transformation $\vec E\to \vec B$, $\vec B\to -\vec E$, but the
action is not. We show that by introducing appropriate auxiliary fields
it is possible to make this duality a manifest symmetry of the action.
Although this process sacrifices manifest Lorentz invariance,
the action is invariant under a certain set of transformations
that reduce to the usual Lorentz transformations when
the auxiliary fields are eliminated by their equations of motion.
We show how to
couple this theory to gravity and to make it supersymmetric
while maintaining
manifest duality symmetry. Generalizations to higher dimensions and
other systems are discussed briefly.

In sect.3 the formalism developed in sect.2 is used to write down an
action that is equivalent to the action of the dimensionally reduced
$D=10$ $N=1$ supergravity theory,
but which has manifest O(6,22) and SL(2,R)
invariance.
In this form of the action, the fields that transform under
SL(2,R) and O(6,22) are treated quite symmetrically.
In this sense, SL(2,R) and O(6,22) appear to be on equal footing.
This action is not manifestly general coordinate invariant, but (as above)
it does have general coordinate invariance nevertheless.

When the auxiliary fields of
the SL(2,R)$\times$O(6,22) invariant action
are eliminated by their equations
of motion,
the original action of ref.\RMAHSCH\ is recovered.
In the special case where the various four-dimensional fields
that originate as components of U(1)$^{16}$ gauge fields in ten
dimensions
are set to zero, there is no preferred choice as to which fields
should be regarded as auxiliary.
In particular, choosing a different set of fields in the manifestly
SL(2,R)$\times$O(6,22) invariant formulation to be the auxiliary fields, and
eliminating them by their equations of motion, gives rise to a manifestly
SL(2,R) and general coordinate invariant formulation of the theory (at
the sacrifice of manifest O(6,22) symmetry).

Although this analysis puts SL(2,R) and O(6,22) symmetry on a very
symmetric footing from the point of view of the four-dimensional effective
field theory, the O(6,22) invariant formulation of the theory could be
regarded as more fundamental, since it is the formulation that appears
naturally in the dimensional reduction of the $N=1$ supergravity theory
from ten to four dimensions. However,
in sect.4 we remove this asymmetry by showing that it is the SL(2,R) and
general coordinate invariant (but not manifestly O(6,22) invariant)
formulation that arises naturally in
the dimensional reduction of the dual formulation of the $N=1$
supergravity theory in ten dimensions based on a six-form potential
with a seven-form field
strength. Since the fields in this dual formulation couple more
naturally to the five-brane\RDUFF, we speculate that the SL(2,Z)
symmetry may have a
more natural realization in the theory of five-branes.
In particular, we show that when expressed in terms of the natural
variables of the five-brane theory, the complex field that transforms
under the SL(2,Z) symmetry takes a form very similar to the fields that
transformed under the target space duality symmetry, expressed in terms
of the natural variables of the string theory.

Sect. 5 gives a summary of our results and some comments.
In particular, we comment on
a possible reformulation of the $N=1$ supergravity action in ten
dimensions, which, upon dimensional reduction, would give the manifestly
SL(2,R)
invariant form of the effective action even when the U(1)$^{16}$ gauge
fields in ten dimensions are included in the theory.

The appendix contains part of the analysis involved in the dimensional
reduction of the dual formulation of $D=10$ $N=1$ supergravity theory.

\chapter{Duality Invariant Einstein--Maxwell Action}

In this section we discuss the construction of an action that is
equivalent to the usual Einstein--Maxwell action, but is manifestly invariant
under a duality symmetry that reduces to the usual $\vec E\to \vec B$,
$\vec B\to -\vec E$ symmetry when the auxiliary fields of the theory are
eliminated by their equations of motion. The method that we use
is very similar to one introduced by Henneaux and Teitelboim
to solve the problem of constructing an action
for the self-dual $(2q+1)$-form field strength in
$4q+2$ dimensions\RHENN. In the special case of two dimensions, it was
discovered independently by Floreanini and Jackiw\RJACKIW\ and used by
Tseytlin for the
construction of a manifestly duality invariant scalar field theory in
two dimensions\RTSEYTLIN. In each of these papers, the key ingredient was
to give up manifest Lorentz invariance of the action.
This will also be the key ingredient in our construction. One of the
main differences between the analysis of the papers mentioned above and
our analysis is the dimensionality of space-time; whereas the analysis of
the previous papers are applicable in 2, 6, 10, $\ldots$ dimansions, our
analysis will be in 4 dimensions. However, at the end of this section we
shall discuss the generalization of our analysis to any dimension. We
also clarify the relationship between our results and those of ref.\RHENN.

This section will be divided into five subsections. In subsection 2.1,
we present an action in four dimensions, which has manifest
duality symmetry and is equivalent to free Maxwell theory. The action
reduces to Maxwell's action when the auxiliary fields are eliminated by their
equations of motion. Although this action is not manifestly Lorentz
invariant, we shall show that the action is, in fact, invariant under a
set of transformations that reduce to the standard Lorentz
transformations when the auxiliary fields are eliminated by their
equations of motion.
In subsection 2.2 we show how to couple this theory to
gravity while preserving manifest duality symmetry.
This gives rise to a theory that is not manifestly invariant under
general coordinate transformations, but is invariant under a set of
transformations that reduce to the usual general coordinate
transformations when the auxiliary fields are eliminated by their
equations of motion. Furthermore, the action reduces to the usual
Maxwell action in curved space-time when the auxiliary
fields are eliminated by their equations of motion.
In subsection 2.3 the construction is generalized
to a field theory of $p$-form fields in
$2p+2$ dimensions for any integer $p$, and the
relationship between our action and that of ref.\RHENN\ is discussed.
In subsection 2.4, the construction
of subsection 2.1 is generalized to the field
theory of $m$-form fields in $d$ dimensions for any $m$ and
$d$, and the action is written in a
form in which the original field, and the dual
$(d-m-2)$-form field appear on an equivalent footing.
Finally, in subsection 2.5 we show how to supersymmetrize
our version of Maxwell's action (as described in subsection 2.1), while
preserving manifest duality symmetry.

\section{Duality Invariant Action}

The basic idea of our construction is to introduce independent gauge
fields for the electromagnetic field strength and its dual.
The fact that the two
field strengths are the duals of one another is then arranged
to be a consequence of the equations of motion. Accordingly,
the basic field variables of our action are a pair of gauge fields
$A_\mu^{(\alpha)}$ ($0\le \mu\le 3$, $1\le\alpha\le 2$). We begin with
flat space-time. The appropriate action is then
$$
S=-{1\over 2}\int d^4x \Big(B^{\a i}\cl_{\alpha\beta} E_i^{(\beta)}
+B^{\a i} B^{\a i}\Big),
\eqn\etwoone
$$
where
$$
E_i^{\a}=\p_0 A_i^{\a}-\p_i A_0^{\a}, \quad B^{\a i}=\epsilon^{ijk} \p_j
A_k^{\a} \quad 1\le i, j, k \le 3
\eqn\etwotwo
$$
and
$$
\cl=\pmatrix{0 & 1\cr -1 & 0\cr}.
\eqn\etwothree
$$
This action has the following gauge invariances
$$
\delta A_0^{\a}=\Psi^{\a}, \quad \delta A_i^{\a}=\p_i \Lambda^{\a}.
\eqn\etwofour
$$
Using the gauge transformation parameter $\Psi^{\a}$, we can set
$$
A_0^{\a}=0.
\eqn\etwofive
$$
Since $A_0^{\a}$ only appears as part of a total derivative in the
action, no equations of motion are lost. (This is to be contrasted with
choosing $A_0=0$ gauge in the usual formulation of Maxwell theory.)
The equation of motion of the field $A_i^{(2)}$ now gives
$$
\epsilon^{ijk}\p_j (B^{(2)k} -E^{(1)}_k)=0.
\eqn\etwosix
$$
Since this does not involve any time derivative of $A^{(2)}_i$, we can
treat $A_i^{(2)}$ as an auxiliary field, and eliminate it from the action
\etwoone\ by using eq.\etwosix.
Eq.\etwosix\ gives
$$
B^{(2)k}=E^{(1)}_k+\p_k\phi
\eqn\etwoseven
$$
for some $\phi$.
Using the freedom associated with the
gauge transformation parameter $\Lambda^{(1)}$, we can set
$\phi=0$, so that
eq.\etwoseven\ reduces to
$$
B^{(2)k}=E^{(1)}_k.
\eqn\etwoeight
$$
Substituting the value of $B^{(2)k}$ given in eq.\etwoeight\ into the
action \etwoone, we get back the usual Maxwell action for the field
$A^{(1)}_\mu$
$$
-{1\over 2} \int d^4 x (B^{(1)i}B^{(1)i} - E^{(1)}_i E^{(1)}_i)
\eqn\etwonine
$$
in the gauge $A^{(1)}_0=0$. The Gauss's law constraint, $\p_i E_i^{(1)}=0$,
is a consequence of the Bianchi identity for $B^{(2)k}$ in eq.
\etwoeight.
Note that \etwoone\ is first order in time derivatives, and therefore
it is
well-suited to a Hamiltonian analysis.

We now return to the original action $S$ given in eq.\etwoone\ and study
its symmetries.
First of all we note that this action is manifestly invariant under the
duality symmetry
$$
A^{\a}_\mu\to \cl_{\alpha\beta} A^{(\beta)}_\mu,
\eqn\etwoten
$$
which implies the transformation
$$
\pmatrix{B^{(1)i}\cr E_i^{(1)}\cr}\to \cl \pmatrix{B^{(1)i}\cr
E^{(1)}_i\cr}
\eqn\etwoeleven
$$
when we use the equation of motion \etwoeight\ of $A_i^{(2)}$.
Note that in the usual formulation of Maxwell's theory, the duality
transformation is a highly non-local transformation on the vector
potential. In contrast, here it is a local transformation on the fields
$A^{\a}_\mu$.

The action given in eq.\etwoone\ is manifestly invariant under rotations,
but not manifestly invariant under Lorentz boosts.
Nevertheless, it can be checked easily that the action is invariant
under the following transformation in the $A^{\a}_0=0$ gauge:
$$
\delta A_i^{\a}= x^0v^k\p_k A_i^{\a}+\vec v.\vec x \cl_{\alpha\beta}
\epsilon^{ijk} \p_j A_k^{(\beta)},
\eqn\etwotwelve
$$
where $\vec v$ is an arbitrary constant three-dimensional vector.
Furthermore, if we use the equations of motion \etwoeight, the
above transformation reduces to
$$
\delta A_i^{(1)}=x^0 v^k\p_k A^{(1)}_i +\vec v.\vec x \p_0 A^{(1)}_i,
\eqn\etwothirteen
$$
which is the usual Lorentz transformation law of the field $A^{(1)}_i$ in
the $A^{(1)}_0=0$ gauge.

\section{Coupling to Gravity}

We shall now generalize the action \etwoone\ to curved space-time in
such a way that when the fields $A^{(2)}_\mu$ are eliminated using their
equations of motion, we recover the Maxwell action for the field
$A^{(1)}_\mu$ in curved space-time
$$
-{1\over 4} \int d^4 x \sqrt{- g} g^{\mu\rho} g^{\nu\sigma}
F^{(1)}_{\mu\nu} F^{(1)}_{\rho\sigma}.
\eqn\etwoeighteen
$$
In order to do this, we start with the most general form of the
action that is first order in time derivatives, invariant
under the duality transformation \etwoten, and invariant under the
gauge transformations \etwofour. This is given by
$$
S_g=-{1\over 2}\int d^4 x\Big[ B^{\a i} \cl_{\alpha\beta} E^{(\beta)}_i
+t_{ij} B^{\a i} B^{\a j} + u_{ij} B^{\a i} \cl_{\alpha\beta}
B^{(\beta)j}\Big].
\eqn\etwofourteena
$$
Here $t_{ij}$ and $u_{ij}$ are unknown coefficients that are determined
by first eliminating the fields $A_i^{(2)}$ from the action
\etwofourteena\ by using their equations of motion, and then demanding that
the resulting action is identical to the action \etwoeighteen.
It turns out that this procedure determines the coefficients $t_{ij}$
and $u_{ij}$
uniquely.
The final action obtained this way is given by
$$
S_g=-{1\over 2}\int d^4 x \Big[ B^{\a i}\cl_{\alpha\beta} E^{(\beta)}_i
-{g_{ij}\over \sqrt{- g} g^{00}} B^{\a i} B^{\a j}+\epsilon^{ijk}
{g^{0k}\over g^{00}} B^{\a i}\cl_{\alpha\beta} B^{(\beta) j}\Big].
\eqn\etwofourteen
$$
Here, as in eq. \etwoeighteen, $\sqrt{-g} = \sqrt{-\det(g_{\mu\nu})}$
and $g^{\mu\nu}$ is the
inverse of $g_{\mu\nu}$, the ordinary four-dimensional metric.
These conventions are retained even when space and time components
are enumerated separately.
By rewriting this formula in terms of $F^{\a}_{ij}$ instead of $B^{\a
k}$ general coordinate invariance in the spatial directions becomes
manifest.

The action
$S_g$ is manifestly invariant under the duality transformation
\etwoten\ and the gauge transformations \etwofour.
Although $S_g$ is not manifestly invariant under general coordinate
transformations, it can be shown to be invariant under the following
transformation:
$$
\delta A_i^{\a}=\xi^j\p_j A_i^{\a}+(\p_i\xi^j) A^{\a}_j
+\xi^0\Big\{-{g_{ij}\over \sqrt{- g} g^{00}}\cl_{\alpha\beta}
B^{(\beta)j}-{g^{0k}\over g^{00}}\epsilon^{ijk} B^{\a j}\Big\}.
\eqn\etwofifteen
$$
To see the connection between this transformation and the usual general
coordinate transformation, we eliminate $A_i^{(2)}$ using its equation
of motion.
In the $A^{\a}_0=0$ gauge, the $A^{(2)}_i$ equation of motion is given
by
$$
\epsilon^{ijk}\p_j\Big\{
E^{(1)}_k+{g_{kj}\over \sqrt{- g} g^{00}} B^{(2)j}+\epsilon^{klm}
{g^{0m}\over g^{00}} B^{(1)l}
\Big\}=0.
\eqn\etwosixteen
$$
Choosing the gauge transformation
parameter $\Lambda^{(1)}$ appropriately, this
equation can be integrated to the form
$$
E^{(1)}_k+{g_{kj}\over \sqrt{- g} g^{00}} B^{(2)j}+\epsilon^{klm}
{g^{0m}\over g^{00}} B^{(1)l}=0.
\eqn\etwoseventeen
$$
If we now substitute the expression for $A_i^{(2)}$ obtained from
eq.\etwoseventeen\ into the expression for $\delta A^{(1)}_i$ given in
eq.\etwofifteen, we get
$$
\delta A_i^{(1)}=\xi^j\p_j A^{(1)}_i +(\p_i \xi^j) A^{(1)}_j
+\xi^0\p_0 A^{(1)}_i.
\eqn\etwonineteen
$$
This is the standard general coordinate transformation law of a vector
under an infinitesimal coordinate transformation $x^\mu\to x^\mu+\xi^\mu$
in the $A^{(1)}_0=0$ gauge.

\section{Generalization to $p$-form Fields in $2p+2$ Dimensions}

In $2p+2$ dimensions, we start with a pair of $p$-form gauge potentials
$A^{\a}_{\mu_1\ldots \mu_p}$ ($0\le \mu_k\le 2p+1$, $1\le\alpha\le 2$),
and define
$$\eqalign{
E^{\a}_{i_1\ldots i_p}=&\p_0 A^{\a}_{i_1\ldots i_p}
-(-1)^p\p_{[i_1}A^{\a}_{i_2\ldots i_p]0},\cr
B^{\a i_1\ldots
i_p}=&{1\over p!}
\epsilon^{i_1\ldots i_p j_1\ldots j_{p+1}}\p_{j_1} A^{\a}_{j_2\ldots
j_{p+1}},} \quad 1\le i_k, j_k\le 2p+1
\eqn\etwotwenty
$$
and
$$
\cl^{(p)}=\pmatrix{0 & 1\cr (-1)^p & 0\cr}.
\eqn\etwotwentyone
$$
In terms of these quantities, the generalization of the action
\etwoone\ is given by
$$
S= -{1\over 2. p!}\int d^{2p+2}x [B^{\a i_1\ldots i_p}
\cl^{(p)}_{\alpha\beta} E^{(\beta)}_{i_1\ldots i_p}
+B^{\a i_1\ldots i_p} B^{\a i_1\ldots i_p}].
\eqn\etwotwentytwo
$$
This action is invariant under the gauge transformations
$$
\delta A^{\a}_{0 i_1\ldots i_{p-1}}=\Psi^{\a}_{i_1\ldots i_{p-1}}, \quad
\delta A^{\a}_{i_1\ldots i_p}=\p_{[i_1}\Lambda^{\a}_{i_2\ldots i_p]},
\eqn\etwotwentythree
$$
the duality transformation
$$
A^{\a}_{\mu_1\ldots \mu_p}\to \cl^{(p)}_{\alpha\beta}
A^{(\beta)}_{\mu_1\ldots \mu_p},
\eqn\etwotwentythreea
$$
and the `Lorentz transformation'
$$
\delta A^{\a}_{i_1\ldots i_p}=x^0 v^j\p_j A^{\a}_{i_1\ldots i_p}
+(-1)^{p+1}
\vec v.\vec x \cl^{(p)}_{\alpha\beta} B^{(\beta)i_1\ldots i_p}.
\eqn\etwotwentysix
$$
Using the gauge transformation parameter $\Psi^{\a}$ we can set the
gauge $A^{\a}_{0i_1\ldots i_{p-1}}=0$.
If we now eliminate the fields $A^{(2)}_{i_1\ldots i_p}$ using their
equations of motion, we recover the standard action for a $(p+1)$-form
field strength in $2p+2$ dimensions
$$
-{1\over 2. (p+1)!}\int d^{2p+2}x F^{(1)}_{\mu_1\ldots \mu_{p+1}}
F^{(1)}_{\nu_1\ldots \nu_{p+1}} \eta^{\mu_1\nu_1} \ldots
\eta^{\mu_{p+1}\nu_{p+1}},
\eqn\etwotwentyfour
$$
where
$$
F^{(1)}_{\mu_1\ldots \mu_{p+1}}=\p_{[\mu_1}
A^{(1)}_{\mu_2\ldots \mu_{p+1}]}.
\eqn\etwotwentyfive
$$
Also, in this case the Lorentz transformation law of $A^{(1)}_{i_1\ldots
i_p}$ takes the standard form in the $A^{(1)}_{0 i_1\ldots i_{p-1}}=0$
gauge.

For $p$ even, the matrix $\cl^{(p)}_{\alpha\beta}$ can be diagonalized
to the form $Diag (1, -1)$. The action \etwotwentytwo\ then
describes the direct sum of two decoupled theories.
One of them is described by the
action of a self-dual $(p+1)$-form field strength as written down in
ref.\RHENN, the other
is described by the action of an anti-self-dual $(p+1)$-form field strength.

Coupling this theory to gravity involves a straightforward
generalization of eq.\etwofourteen.
The corresponding action is given by
$$\eqalign{
S_g=& -{1\over 2. p!}\int d^{2p+2}x \Big[ B^{\a i_1\ldots i_p}
\cl^{(p)}_{\alpha\beta} E^{(\beta)}_{i_1\ldots i_p}
-{g_{i_1j_1}\ldots g_{i_pj_p}\over\sqrt{- g} g^{00}} B^{\a i_1\ldots
i_p}B^{\a j_1\ldots j_p}\cr
&+{1\over p!}\epsilon^{i_1\ldots i_p j_1\ldots j_p k} {g^{0k}\over
g^{00}} B^{\a i_1\ldots i_p} \cl^{(p)}_{\alpha\beta} B^{(\beta)
j_1\ldots j_p}\Big].\cr
}
\eqn\etwotwentyseven
$$
This is invariant under the `general coordinate transformation'
$$\eqalign{
\delta A^{\a}_{i_1\ldots i_p}=& \xi^j\p_j A^{\a}_{i_1\ldots i_p}
+(-1)^{p-1}(\p_{[i_1}\xi^j)A^{\a}_{i_2\ldots i_p]j}\cr
&+\xi^0\Big[ (-1)^p{g_{i_1j_1}\ldots g_{i_pj_p}\over
\sqrt{- g} g^{00}}
\cl^{(p)}_{\alpha\beta} B^{(\beta)j_1\ldots j_p} - {g^{0k}\over
g^{00}}\p_{[k}A^{\a}_{i_1\ldots i_p]}\Big]\cr
}
\eqn\etwotwentyeight
$$
If we eliminate $A^{(2)}_{i_1\ldots i_p}$ from the action
\etwotwentyseven\ by its equation of motion, we get back the
covariantized form of the action \etwotwentyfour.
Also, in this case the general coordinate transformation law of the field
$A^{(1)}_{i_1\ldots i_p}$ reduces to the standard form in the $A^{(1)}_{0
i_1\ldots i_{p-1}}=0$ gauge.
Finally, if we diagonalize the matrix $\cl^{(p)}$, we get back the sum
of the action of a self-dual $(p+1)$-form field strength and an
anti-self-dual $(p+1)$-form field strength in curved space-time, as
written down in ref.\RHENN.

\section{$m$-form Fields in $d$ Dimensions}

Let us consider next the free field theory of
an $m$-form field $A_{\mu_1\ldots\mu_m}$ in $d$
dimensions.
The corresponding field strength is
$$
F_{\mu_1\ldots \mu_{m+1}}=\p_{[\mu_1} A_{\mu_2\ldots \mu_{m+1}]}, \quad
0\le \mu_l\le d-1.
\eqn\etenone
$$
The equations of motion and the Bianchi identities are
$$
\eta^{\mu_1\rho_1}\p_{\rho_1}F_{\mu_1\ldots \mu_{m+1}}=0, \quad
\varepsilon^{\mu_1\ldots \mu_{m+2}\nu_1\ldots \nu_{d-m-2}}
\p_{\mu_1} F_{\mu_2\ldots \mu_{m+2}}=0.
\eqn\etentwo
$$
We can dualize this theory by introducing a dual $(d-2-m)$-form
potential,
$B_{\nu_1\ldots \nu_{d-m-2}}$, and the corresponding field strength,
$$
G_{\nu_1\ldots \nu_{d-m-1}}=\p_{[\nu_1} B_{\nu_2\ldots \nu_{d-m-1]}},
\eqn\etenthree
$$
such that
$$
\eta^{\mu_1\rho_1}\ldots \eta^{\mu_{m+1}\rho_{m+1}} F_{\rho_1\ldots
\rho_{m+1}}=
{1\over (d-m-1)!}
\varepsilon^{\mu_1\ldots \mu_{m+1}\nu_1\ldots \nu_{d-m-1}}
G_{\nu_1\ldots \nu_{d-m-1}}.
\eqn\etenfour
$$
It is easy to check that the equations of motion of $F$ correspond to
Bianchi identities of $G$ and vice versa.
Examples of such pairs of dual fields are a scalar and a two-form field in
four dimensions, a two-form field and a six-form field in ten dimensions,
etc.

Normally, the action of such a theory is written either in terms of the
original field $A$ or the dual field $B$, but not both. We shall now
write down a form of the action in which $A$ and $B$ appear on an equal
footing. Consider the action
$$\eqalign{
S_0 =& {1\over m! (d-m-1)!} \epsilon^{i_1\ldots i_m j_1\ldots
j_{d-m-1}} F_{0 i_1\ldots i_m} G_{j_1\ldots j_{d-m-1}}\cr
&+{1\over 2\cdot (m+1)!}F_{i_1\ldots i_{m+1}} F_{i_1\ldots i_{m+1}}\cr
&+{1\over 2\cdot (d-m-1)!}
G_{i_1\ldots i_{d-m-1}} G_{i_1\ldots i_{d-m-1}},\cr
&\qquad\qquad\qquad\qquad 1\le i_l, j_l\le d.\cr
}
\eqn\etenfive
$$
The action \etenfive\ is invariant under the following gauge
transformations:
$$\eqalign{
\delta A_{0 i_1\ldots i_{m-1}}=&\Psi^{(1)}_{i_1\ldots i_{m-1}}, \quad
\delta B_{0 i_1\ldots i_{d-m-3}}=\Psi^{(2)}_{i_1\ldots i_{d-m-3}}\cr
\delta A_{i_1\ldots i_m}=&\p_{[i_1}\Lambda^{(1)}_{i_2\ldots i_m]}, \quad
\delta B_{i_1\ldots i_{d-m-2}}=\p_{[i_1}\Lambda^{(2)}_{i_2\ldots
i_{d-m-2}]}\cr
}
\eqn\etenseven
$$
Using the gauge transformation parameters $\Psi^{(1)}$, $\Psi^{(2)}$ we
can set the gauge
$$
A_{0 i_1\ldots i_{m-1}}=0, \quad B_{0 i_1\ldots i_{d-m-3}}=0.
\eqn\eteneight
$$
Finally, with the help of the gauge transformation parameters
$\Lambda^{(1)}$ and $\Lambda^{(2)}$, the equations of motion derived
from the action \etenfive\ can be shown to be precisely those given in
eq.\etenfour.
Also, if we eliminate either the $A$ or the $B$ fields from the action
\etenfive\ by their equations of motion, we get back the standard free
action for the other field.

In many cases, the free equations \etenfour\ get modified by the
addition of a Chern-Simons term to the field strength.
The duality relations \etenfour\ then get modified to
$$
\eta^{\mu_1\rho_1}\ldots \eta^{\mu_{m+1}\rho_{m+1}} (F_{\rho_1\ldots
\rho_{m+1}}+\Omega_{\rho_1\ldots \rho_{m+1}})
={1\over (d-m-1)!}
\varepsilon^{\mu_1\ldots \mu_{m+1}\nu_1\ldots \nu_{d-m-1}}
G_{\nu_1\ldots \nu_{d-m-1}},
\eqn\etennine
$$
where $\Omega$ is an $m+1$ form.\foot{Note that although the
addition of $\Omega$ to $F$ seems to destroy the symmetry between $F$ and
$G$, we could have added the dual of $\Omega$ to $G$ with the same
effect.}
We now ask the question: Is it possible to modify the action \etenfive\
in such a way that the corresponding equations of motion are the
modified eqs.\etennine?
The answer to this question is yes. We simply need to add the term
$$
\eqalign{S_1= & {1\over (m+1)!}F_{i_1\ldots i_{m+1}}\Omega_{i_1\ldots
i_{m+1}}\cr
& +{1\over m!}{1\over (d-m-1)!} \epsilon^{i_1\ldots i_m j_1\ldots
j_{d-m-1}}\Omega_{0 i_1\ldots i_m} G_{j_1\ldots j_{d-m-1}} \cr}
\eqn\etenten
$$
to the action $S_0$ given in eq.\etenfive.

\section{Supersymmetrization of the Duality-Invariant Maxwell Action}

We shall now discuss how to supersymmetrize the duality-invariant
Maxwell action while
preserving manifest duality invariance. Since we are using a formalism
that is not manifestly Lorentz invariant, we can use two-component
spinors instead of four-component spinors for describing the fermionic
fields in this theory. We know that the supersymmetry partner of a
vector field in four dimensions should be a Majorana spinor. Such a
field can be represented by a pair of complex two-component
spinors $\psi^{\a}$ ($1\le\alpha\le 2$) satisfying the condition
\foot{These are essentially the same thing as what is often described as
two-component spinors with dotted and undotted indices. The notation
used here is much more natural in the present context.}
$$
\psi^{\a *}=\sigma_2\cl_{\alpha\beta}\psi^{(\beta)}.
\eqn\estwo
$$
Here $\sigma_i$ are the standard Pauli matrices. They act on
the implicit spinor index of $\psi^{\a}$.
The full action is now given by
$$
\eqalign{S = \int d^4 x \big[ & -{1\over 2}
(B^{\a i}\cl_{\alpha\beta} E_i^{(\beta)}
+B^{\a i} B^{\a i})\cr
& + i\psi^{\a\dagger}\p_0\psi^{\a} -\psi^{\a\dagger}
\cl_{\alpha\beta} \sigma_k\p_k\psi^{(\beta)}\big].\cr}
\eqn\esone
$$
This action is invariant under the following supersymmetry
transformations:
$$\eqalign{
\delta\psi^{\a} =& {1\over 2} (\cl_{\alpha\beta}\sigma_k
B^{(\beta)k}\epsilon -\sigma_k B^{\a k}\sigma_2\epsilon^*)\cr
\delta A_i^{\a} =& i\psi^{\a \dagger}\sigma_i\epsilon -
i\psi^{(\beta)\dagger}\cl_{\alpha\beta}\sigma_i\sigma_2\epsilon^*,\cr
}
\eqn\esthree
$$
where $\epsilon$ is an arbitrary two-component complex spinor.

In order to see that the action \esone\ and the transformation laws
\esthree\ reduce to the standard action and
supersymmetry transformation laws in four dimensions
when we eliminate the auxiliary fields
$A_i^{(2)}$ by their equations of motion, we introduce four-component
spinors
$$
\psi=\pmatrix{\psi^{(1)}\cr \psi^{(2)}\cr}
\eqn\esfour
$$
$$
\eta=i\pmatrix{\sigma_2\epsilon^*\cr \epsilon\cr}
\eqn\esseven
$$
and the four-dimensional matrices $\gamma^\mu$ such that
$$
\gamma^0\gamma^i=\pmatrix{0 & -i\sigma_i\cr i\sigma_i & 0\cr}.
\eqn\esfive
$$
In terms of these quantities, the fermion bilinear term in eq.\esone\
may be written as
$$
-i\bar\psi\gamma\cdot\partial \psi.
\eqn\essix
$$
Also, using eq.\etwoeight\ the supersymmetry transformation laws given
in eq.\esthree\ may be rewritten as
$$
\delta A^{(1)}_i=i\bar\psi\gamma^i\eta, \quad \delta\psi={1\over
4}\gamma_\mu\gamma_\nu F^{\mu\nu}\eta,
\eqn\eseight
$$
which are the standard supersymmetry transformation laws in four
dimensions.
Finally, from eqs.\estwo, \esfour\ and \esseven\ we see that
$\psi$ and $\eta$ satisfy the Majorana condition
$$
\psi^*=i\gamma^0\gamma^2\psi, \quad \eta^*=i\gamma^0\gamma^2\eta.
\eqn\esnine
$$

Since the fermi terms in eq.\esone\ have been shown to agree with the
standard formula for the kinetic term of a spinor, they can be coupled
to gravity, thereby achieving general coordinate invariance and local Lorentz
invariance, in the standard way, namely
$$ \int d^4 x \sqrt{-g}\big[ i e_0^{\mu}\psi^{\a\dagger}D_{\mu}
\psi^{\a} - e_k^{\mu}\psi^{\a\dagger}
\cl_{\alpha\beta} \sigma_kD_{\mu}\psi^{(\beta)}\big].
\eqn\esten
$$
where $D_\mu$ denotes the covariant derivative involving the spin
connection.
The coupling to supergravity can then be worked out by standard methods.

\chapter{ Low-Energy Effective Action in
String Theory with Manifest SL(2,R) Symmetry}

The low-energy effective action describing heterotic string theory
compactified on a six-dimensional torus at a generic point in the moduli
space is given by\RMAHSCH\RDYONO\
$$\eqalign{
\int & d^4 x\sqrt{- g} \Big[ R -{1\over 2(\lambda_2)^2}g^{\mu\nu}
\p_\mu\lambda \p_\nu\bar\lambda -{\lambda_2\over 4}F^a_{\mu\nu}
(LML)_{ab} F^{b\mu\nu}\cr
& +{\lambda_1\over 4} F^a_{\mu\nu} L_{ab} \tilde F^{b\mu\nu}
+{1\over 8} g^{\mu\nu} Tr(\p_\mu M L \p_\nu M L)\Big],\cr
}
\eqn\ethreeone
$$
where $A^a_\mu$ ($1\le a\le 28$) are a set of 28 abelian gauge fields
and
$$
F^a_{\mu\nu}=\p_\mu A^a_\nu -\p_\nu A^a_\mu, \quad \tilde F^{a\mu\nu}
={1\over 2} (\sqrt{- g})^{-1}\epsilon^{\mu\nu\rho\sigma}
F^a_{\rho\sigma}.
\eqn\ethreefour
$$
$$
\lambda=\lambda_1+i\lambda_2
\eqn\ethreetwo
$$
is a complex scalar field,
$$
L=\pmatrix{ 0 & I_6 & 0\cr I_6 & 0 & 0\cr 0 & 0 & -I_{16}},
\eqn\ethreethree
$$
and $M$ is a 28$\times$28 matrix-valued scalar field satisfying the
constraints
$$
M^T=M, \quad M^T L M=L.
\eqn\ethreefive
$$
The action \ethreeone\ is manifestly invariant under an O(6,22)
transformation
$$
M\to \Omega^T M \Omega, \quad A^a_\mu\to \Omega^T_{ab} A^b_\mu
\eqn\ethreesix
$$
where $\Omega$ is a $28\times 28$ matrix satisfying
$$
\Omega^T L \Omega = L.
\eqn\ethreesix
$$
The equations of motion derived from the action \ethreeone\ have a
further SL(2,R) symmetry\ROLD\RTSW\RDUALITY\RSCHWARZ, given by
$$
\lambda\to {a\lambda + b\over c\lambda+d}, \quad F^a_{\mu\nu}
\to c\lambda_2 (ML)_{ab}\tilde F^b_{\mu\nu} +(c\lambda_1+d)
F^a_{\mu\nu},
\quad ad - bc=1.
\eqn\ethreeseven
$$
The action \ethreeone, however, is not invariant under this SL(2,R)
transformation. More specifically, the terms involving the gauge fields
are not invariant; the other terms are invariant.

In subsection (3.1) we shall show that, using the formalism of the previous
section, we can write down a manifestly SL(2,R)$\times$O(6,22)
invariant action, which is equivalent to the action \ethreeone. The price
that we'll have to pay is again manifest general coordinate invariance
of the action. We shall also see that SL(2,R)
and O(6,22) transformations appear in a symmetric manner
in the resulting action.
The analysis of this subsection raises
the question whether it is possible to write down a third form of
the action in which SL(2,R) and general coordinate invariance of the
action are manifest, but O(6,22) appears only as a symmetry of
the equations of motion. In subsection (3.2) we show that this is
possible for a restricted class of field configurations---the
configurations for which all four-dimensional fields arising out of
dimensional reduction of ten dimensional U(1)$^{16}$ gauge
fields are set to zero.

\section{Manifestly SL(2,R)$\times$O(6,22) invariant action}

We shall carry out the construction of a manifestly SL(2,R)$\times$O(6,22)
invariant action in three steps. In the first step we shall show how to
generalize the action \etwoone\ to the case of multicomponent gauge
fields. As we shall see, this will automatically introduce the matrix
$M$ appearing in eq.\ethreeone\ and satisfying \ethreefive\
into the action. In the second step, we
shall show how to couple the action \etwoone\ to the complex field
$\lambda$ transforming as in eq.\ethreeseven\ in an SL(2,R) invariant
manner. Finally, in the third step, we shall combine steps 1 and 2, as
well as the result of the last section, to couple the gauge fields to the
matrix-valued field $M$, the complex field $\lambda$, and the metric
$g_{\mu\nu}$ in an SL(2,R) invariant fashion.

\noindent Step 1. We consider generalization of the action \etwoone\ to
multicomponent gauge fields $A_\mu^{(a,\alpha)}$.
The general form of the action consistent with the requirement of
duality symmetry \etwoten, gauge invariance \etwofour, rotational
invariance, and invariance under the parity transformation
$A_i^{(a,\alpha)}(x^0, \vec x)\to (-1)^\alpha
A_i^{(a,\alpha)}(x^0, -\vec x)$,
is given by\foot{We assume that the matrices $P$, $Q$ etc. are inert
under these symmetries. Otherwise more general possibilities may
arise.}
$$
S_{P,Q}=-{1\over 2}\int d^4x[B^{\aa i}\cl_{\alpha\beta} Q_{ab} E_i^{\bb} +
B^{\aa i} P_{ab} B^{(b,\alpha) i}],
\eqn\ethreefourteen
$$
where $Q$ is a space-time independent matrix, $P$ is a space-time
dependent matrix (in general), and
$$
B^{\aa i}=\epsilon^{ijk}\p_j A^{\aa}_k, \quad
E^{\aa}_i= \p_0 A^{\aa}_i-\p_i A^{\aa}_0.
\eqn\ethreefifteen
$$
Since only the symmetric parts of $Q$ and $P$ contribute to the action,
we can choose these matrices to be symmetric without any loss of
generality.
Also, using the freedom of a linear redefinition of the gauge fields,
$A^{\aa}_i\to S_{ab}A^{(b,\alpha)}_i$, where $S$ is a space-time
independent matrix, we can ensure that the matrix $Q$ has eigenvalues
$\pm 1$, so that $Q^2=I$.
If we now eliminate the fields $A_i^{(a,2)}$ from the action
\ethreefourteen\ using their equations of motion, we get the action
$$
-{1\over 2}\int d^4 x[B^{(a,1)i}P_{ab}B^{(b,1)i} - E^{(a,1)}_i (Q P^{-1}
Q)_{ab}E^{(b,1)}_i].
\eqn\ethreesixteen
$$
This action is manifestly Lorentz invariant provided
$$
QP^{-1}Q=P.
\eqn\ethreeseventeen
$$
Comparing eqs.\ethreesixteen, \ethreeseventeen, with \ethreeone,
\ethreefive\ in the background $g_{\mu\nu}=\eta_{\mu\nu}$, $\lambda=i$,
we see that we need the identification
$$
Q=L, \qquad P=LML.
\eqn\ethreeeighteen
$$
The action \ethreefourteen\ is not manifestly Lorentz invariant.
But it is invariant under hidden Lorentz transformations, which are direct
generalizations of the Lorentz transformation laws \etwotwelve.
Since these transformation laws can always be derived from the general
coordinate transformation laws of the final action that we shall write
down, we shall not write down the Lorentz transformation laws of the
fields $A^{\aa}_i$ explicitly here.

\noindent Step 2. We now go back to the action \etwoone\ and try to
couple the complex field $\lambda$ to this action in an SL(2,R)
invariant fashion. In order to do this, we first introduce a matrix
$$
\cm(\lambda) ={1\over \lambda_2}\pmatrix{1 & \lambda_1\cr \lambda_1 &
|\lambda|^2\cr},
\eqn\ethreeeight
$$
satisfying,
$$
\cm^T=\cm, \quad \cm\cl\cm^T=\cl
\eqn\enewone
$$
Under the SL(2,R) transformation \ethreeseven\ of the field $\lambda$,
the matrix $\cm$ transforms in a simple manner,
$$
\cm\to \omega^T\cm\omega,
\eqn\ethreenine
$$
where
$$
\omega=\pmatrix{d & b\cr c & a\cr}.
\eqn\ethreeninea
$$
An SL(2,R) invariant coupling of the action \etwoone\ to the field
$\lambda$ may now be written down as follows:
$$
S_\lambda=-{1\over 2}\int d^4 x[B^{\a i}\cl_{\alpha\beta} E^{(\beta)}_i
+B^{\a i} (\cl^T\cm \cl)_{\alpha\beta}B^{(\beta) i}].
\eqn\ethreeten
$$
Using the relation
$$
\omega \cl \omega^T=\cl,
\eqn\ethreetwelve
$$
one can easily see that the action \ethreeten\ is invariant under
the transformation \ethreenine\ on $\cm$, together with the
transformation
$$
A^{\a}_i\to (\omega^T)_{\alpha\beta} A^{(\beta)}_i.
\eqn\ethreeeleven
$$
After eliminating the fields $A^{(2)}_i$ using their equations of
motion, we get the action
$$
-{1\over 4}\int d^4 x  (\lambda_2 F^{(1)}_{\mu\nu} F^{(1)}_{\rho\sigma}
-\lambda_1 F^{(1)}_{\mu\nu} \tilde F^{(1)}_{\rho\sigma})\eta^{\mu\rho}
\eta^{\nu\sigma}.
\eqn\ethreethirteen
$$
The gauge field dependent part of the action
\ethreeone\ in flat background, and  for $M=I$, $L=I$, is precisely $22$
copies of this action.
Also, the duality transformation \ethreeeleven\ takes precisely the
form of eq.\ethreeseven\ with $M=L=I$ after we eliminate $A^{(2)}_i$
from these transformation laws using their equations of motion.
The gauge fields $A^{(2)}_\mu$ may be identified with the dual
vector potentials introduced by
Kallosh and Ortin\RKALORTTWO.

Again, the action \ethreeten\ has a hidden Lorentz invariance.
But we shall not write down the Lorentz transformation laws of the gauge
fields explicitly here.

\noindent Step 3. We shall now combine eqs.\etwofourteen, \ethreeten\
and \ethreefourteen\ together, for the identification given in
eq.\ethreeeighteen, to obtain the manifestly SL(2,R) invariant
coupling of the gauge fields $A^{\aa}_\mu$ to the fields $M$, $\lambda$
and $g_{\mu\nu}$.
The resulting action is
$$\eqalign{
S_{\lambda, M, g}=-{1\over 2}\int d^4 x\Big[& B^{\aa i}\cl_{\alpha\beta}
L_{ab} E_i^{\bb}
+ \epsilon^{ijk}{g^{0k}\over g^{00}} B^{\aa i}\cl_{\alpha\beta} L_{ab}
B^{\bb j}\cr
&
-{g_{ij}\over \sqrt{- g} g^{00}} B^{\aa i}
(\cl^T\cm\cl)_{\alpha\beta} (LML)_{ab} B^{\bb j}\Big].\cr}
\eqn\ethreenineteen
$$
If we eliminate the fields $A^{(a,2)}_i$ from this action using their
equations of motion, we get back the gauge field dependent part of the
action \ethreeone\
$$
-{1\over 4}\int d^4 x \sqrt{- g} [\lambda_2 F^{(a,1)}_{\mu\nu}
(LML)_{ab} F^{(b,1)\mu\nu} -\lambda_1 F^{(a,1)}_{\mu\nu} L_{ab} \tilde
F^{(b,1)\mu\nu}].
\eqn\ethreetwentyone
$$

The action \ethreenineteen\ is manifestly invariant under the O(6,22)
transformation given in eqs.\ethreesix\ and the SL(2,R) transformation given
in eqs.\ethreenine, \ethreeeleven. It is not manifestly invariant under
general coordinate transformations. However, it can be checked
that it is invariant under the transformation
$$
\eqalign{\delta A_i^{\aa}=&\xi^j\p_j A_i^{\aa} +(\p_i \xi^j)
A_j^{\aa}\cr
&-\xi^0 \Big\{ {g_{ij}\over \sqrt{- g}
g^{00}}(\cm\cl)_{\alpha\beta} (ML)_{ab} B^{\bb j}+{g^{0k}\over g^{00}}
\epsilon^{ijk} B^{\aa j}\Big\},\cr}
\eqn\ethreetwenty
$$
which generalizes \etwofifteen\ and
reduces to the usual general coordinate transformation law of the
field $A^{(a,1)}_i$ in the $A^{(a,1)}_0=0$ gauge when the fields
$A^{(a,2)}_i$ are eliminated by their equations of motion.

In terms of the matrix $\cm$, the $\lambda$ field kinetic term appearing
in eq.\ethreeone\ can also be written in a manifestly SL(2,R) invariant
form:
$$
{1\over 2(\lambda_2)^2}g^{\mu\nu}\p_\mu\lambda \p_\nu\bar\lambda
={1\over 4} g^{\mu\nu} tr(\p_\mu \cm \cl \p_\nu \cm \cl).
\eqn\ethreetwentyonea
$$
Thus the full action \ethreeone\ may be replaced by
$$\eqalign{
S = \int  d^4 x \Big[& \sqrt{- g}\big\{ R - {1\over 4} g^{\mu\nu}
tr(\p_\mu \cm \cl \p_\nu\cm \cl) + {1\over 8} g^{\mu\nu} Tr(\p_\mu M L
\p_\nu M L)\big\}\cr
&-{1\over 2}\Big\{ B^{\aa i}\cl_{\alpha \beta} L_{ab} E_i^{\bb}
+ \epsilon^{ijk}{g^{0k}\over g^{00}} B^{\aa i}\cl_{\alpha\beta} L_{ab}
B^{\bb j}\cr
&
-{g_{ij}\over\sqrt{- g} g^{00}} B^{\aa i}(\cl^T\cm\cl)_{\alpha\beta}
(LML)_{ab} B^{\bb j} \Big\}\Big]\cr}
\eqn\ethreetwentytwo
$$
In the above equation $Tr$ denotes trace over the indices $a,b$ and $tr$
denotes trace over the indices $\alpha, \beta$.
Note that the matrices $\cm, \cl$ and $M,L$ appear quite
symmetrically in the expression for $S$.

In three dimensions both SL(2,R)
and O(6,22) become part of a larger symmetry group O(8,24)\RMARSCH.
This provides further evidence that SL(2,R) and O(6,22) should play
identical roles in the full string theory.

\section{Action With Manifest SL(2,R) and General Coordinate Invariance}

We have seen that starting with the action \ethreetwentytwo\ and
eliminating the auxiliary fields $A^{(a, 2)}_i$ by their equations of
motion gave the manifestly $O(6,22)$ and general coordinate invariant
action \ethreeone. Note, however, that in the action \ethreetwentytwo\
the various fields $A^{\aa}_\mu$ appear symmetrically, and hence it
is a matter of choice which subset of these fields we treat as
auxiliary fields. If we choose the subset of auxiliary fields to be
invariant under O(6,22) transformations, then we would expect the
final action to be manifestly invariant under O(6,22) transformations, as
was the case in going from the action \ethreetwentytwo\ to \ethreeone.
But the same argument shows that if we choose the set of auxiliary
fields in such a way that the set is invariant under SL(2,R)
transformations, then the resulting action should be manifestly SL(2,R)
invariant, but not manifestly O(6,22) invariant. This naturally gives
rise to the question as to whether it is possible to get a manifestly
SL(2,R) and general coordinate invariant action following this
procedure.

We shall now show that it is possible to obtain such an action provided
we set all the fields arising from the dimensional reduction of
ten-dimensional gauge fields to zero.
In terms of the fields appearing in eq.\ethreeone\ this means that we
now take the gauge fields to have 12 components instead of 28 components,
$L$ to be the 12$\times$12 matrix
$$
L=\pmatrix{0 & I_6\cr I_6 & 0\cr},
\eqn\efourone
$$
and $M$ to be a 12$\times$12 matrix-valued field satisfying the same
constraints \ethreefive\ with respect to the new $L$.
Such a matrix $M$ can be parametrized as
$$
M=\pmatrix{\hG^{-1} & \hG^{-1}\hB\cr -\hB\hG^{-1} & \hG -\hB\hG^{-1}
\hB\cr},
\eqn\efourtwo
$$
where $\hG$ and $\hB$ are $6\times 6$ symmetric and antisymmetric
matrices, respectively, which can be identified with the internal
components of the ten-dimensional metric and antisymmetric tensor
fields, respectively.
The O(6,6)$\times$SL(2,R) invariant form of the action is given by
eq.\ethreetwentytwo\ with the indices $a,b$ running from 1 to 12.

We now start from eq.\ethreetwentytwo\ and eliminate the fields
$A^{(m+6,\alpha)}_i$ ($1\le m\le 6$, $1\le \alpha\le 2$) by their
equations of motion.\foot{Note that this is an SL(2,R) invariant set.}
With appropriate choice of gauge, these equations can be brought to the
form:
$$\eqalign{
g_{ij}B^{(m+6,\alpha)j}=&-\sqrt{-
g}g^{00}\hG_{mn}(\cm\cl^T)_{\alpha\beta}\Big\{ E^{(n,\beta)}_i
+\epsilon^{ijk} {g^{0k}\over g^{00}} B^{(n,\beta) j}\Big\}\cr
& -g_{ij}\hB_{mn} B^{(n,\alpha)j}.\cr}
\eqn\efourfour
$$
Here $i,j,k$ are spatial indices, and $m,n$ are indices denoting the six
internal directions.
If we now substitute this back into the action \ethreetwentytwo, we get
an action of the form:
$$\eqalign{
&\int d^4 x\sqrt{- g}\Big[ R -{1\over 4} g^{\mu\nu} tr(\p_\mu\cm\cl
\p_\nu\cm \cl)+{1\over 8} g^{\mu\nu}Tr(\p_\mu M L \p_\nu M L)\cr
& -{1\over 4}F^{(m,\alpha)}_{\mu\nu}\hG_{mn} (\cl^T\cm\cl)_{\alpha\beta}
F^{(n,\beta)}_{\rho\sigma}g^{\mu\rho}g^{\nu\sigma}
-{1\over 4} F^{(m,\alpha)}_{\mu\nu}\hB_{mn}\cl_{\alpha\beta}
\tilde F^{(n,\beta)}_{\rho\sigma} g^{\mu\rho} g^{\nu\sigma}\Big].\cr
}
\eqn\efourfive
$$
This action is manifestly SL(2,R) and Lorentz invariant, but not
manifestly O(6,6) invariant.
However, since the equations of motion derived from this action are
identical to those derived from the action \ethreetwentytwo, we can
conclude that these equations of motion are also O(6,6) invariant.

\chapter{Manifestly SL(2,R) Invariant Effective Action from Dimensional
Reduction of $N=1$ $D=10$ Supergravity Theory}

In the previous section
we have given a formulation of the low-energy
effective action in heterotic string theory that is manifestly SL(2,R)
and O(6,22) invariant, but not manifestly general coordinate invariant.
We have also shown that in the special case where all the components of
the ten-dimensional gauge fields are set to zero, we can get a
manifestly SL(2,R) and general coordinate invariant action by
sacrificing O(6,22) invariance. This analysis puts the O(6,22) and
SL(2,R) symmetry on an equal footing from the point of view
of the four-dimensional effective field theory. However, it is the
manifestly O(6,22) and general coordinate invariant form of the action
that arises naturally in the dimensional reduction of the $N=1$
supergravity theory in ten dimensions. From this point of view, the
O(6,22) symmetry of the action might appear to be more fundamental than the
SL(2,R) symmetry. In this section we shall get rid of
this asymmetry by showing that it is the
SL(2,R) invariant action \efourfive\
that arises naturally in the dimensional
reduction of another ten-dimensional theory---the dual formulation of
the $N=1$ $D=10$ supergravity theory\RDUALSUGRA.

Before we can write down the field content and action of this ten-dimensional
theory, we must describe our notation. We shall denote
ten-dimensional coordinates by $z^M$ ($0\le M\le 9$), whereas $y^m$ ($4\le
m\le 9$) and $x^\mu$ ($0\le\mu\le 3$) will denote the internal and
space-time coordinates,
respectively. The superscript $^{(10)}$ will denote fields that appear
naturally in the ten-dimensional theory; the fields
which are more natural from the point of
view of four-dimensional theory will not carry this superscript.
The subscript $_S$ will denote
the metric which couples naturally to the string (the one that appears
in the world-sheet action).
Finally, since the fields appearing in the dual formulation
of the $N=1$ $D=10$ supergravity theory couple naturally to the
five-brane\RSTROM\RDUFF,
it is also convenient to introduce a
new metric
that couples naturally to the five-brane; we shall denote this one by the
subscript $_F$.

In the absence of ten-dimensional gauge fields, the only bosonic fields
in the dual formulation of the $N=1$ supergravity theory in 10
dimensions are the metric $G^{(10)}_{FMN}$, the dilaton $\Phi^{(10)}$
and the 6-form field $\ca^{(10)}_{M_1\ldots M_6}$. The bosonic part of
the action may be written as\RDUFFLU\
$$\eqalign{
S=\int d^{10}z& \sqrt{-\det G^{(10)}_F} e^{\Phi^{(10)}/3} (R^{(10)}_F\cr
&-{1\over 2. 7!} G^{(10)M_1N_1}_F \ldots G^{(10)M_7N_7}_F
K^{(10)}_{M_1\ldots M_7} K^{(10)}_{N_1\ldots N_7}),\cr}
\eqn\efiveone
$$
where
$$
K^{(10)}_{M_1\ldots M_7}=\p_{[M_1}\ca^{(10)}_{M_2\ldots M_7]}.
\eqn\efivetwo
$$
The string metric $G^{(10)}_{SMN}$ is related to the five-brane metric
metric $G^{(10)}_{FMN}$ through the relation\RDUFFLU\
$$
G^{(10)}_{FMN}=e^{-\Phi^{(10)}/3}  G^{(10)}_{SMN}.
\eqn\efivethree
$$
In terms of the metric $G_{SMN}^{(10)}$, the first term in the action
\efiveone\ may be written as\RDUFFLU\
$$\eqalign{
S_1\equiv & \int d^{10}z \sqrt{-\det G^{(10)}_F} e^{\Phi^{(10)}/3}
R^{(10)}_F\cr
=&\int d^{10}z \sqrt{-\det G^{(10)}_S} e^{-\Phi^{(10)}} (R^{(10)}_S +
G^{(10)MN}_S \p_M\Phi^{(10)} \p_N\Phi^{(10)}).\cr
}
\eqn\efivefour
$$
Dimensional reduction of this term to four dimensions was already
carried out in ref.\RMAHSCH, so we just state the results here.
We define\RMAHSCH\RDUALITY\
$$\eqalign{
\hG_{mn}= & G^{(10)}_{Smn}, \quad C^m_\mu=\hG^{mn}G^{(10)}_{Sn\mu},
\quad G_{S\mu\nu}=G^{(10)}_{S\mu\nu}-G^{(10)}_{Sm\mu}
G^{(10)}_{Sn\nu}\hG^{mn}\cr
\Phi= & \Phi^{(10)}-{1\over 2}\ln\det\hG, \quad \lambda_2=e^{-\Phi}, \quad
g_{\mu\nu}=e^{-\Phi} G_{S\mu\nu}. \cr
}
\eqn\efivefive
$$
In the above equations $\hG^{mn}$ denotes the matrix inverse of
$\hG_{mn}$.
If we take the various fields to be independent of the internal
coordinates, and normalize $\int d^6 y$ to 1, we get the following form
of the dimensionally reduced action:
$$\eqalign{
S_1 = & \int d^4 x \sqrt{- g} [R - {1\over 2(\lambda_2)^2}
g^{\mu\nu}\p_\mu\lambda_2 \p_\nu\lambda_2 +{1\over 4} g^{\mu\nu}
Tr(\p_\mu \hG \p_\nu \hG^{-1})\cr
& -{1\over 4}\lambda_2\hG_{mn} g^{\mu\rho} g^{\nu\sigma}
F^{(C)m}_{\mu\nu} F^{(C)n}_{\rho\sigma}],\cr
}
\eqn\efivesix
$$
where
$$
F^{(C)m}_{\mu\nu}=\p_\mu C^m_\nu -\p_\nu C^m_\mu.
\eqn\efiveseven
$$

We now need to carry out
the dimensional reduction of the second term in the
action \efiveone.
This has been carried out in detail in
the appendix; here we only quote the result. The final result agrees with
the action \efourfive, provided we make the identifications
$$\eqalign{
\lambda_1 =\ & {1\over 6!}\epsilon^{m_1\ldots m_6} \ca^{(10)}_{m_1\ldots
m_6}\cr
A^{(m,1)}_\mu =\ & C^m_\mu\cr
A^{(m,2)}_\mu =\ &
{1\over 5!} \epsilon^{m m_2\ldots m_6} (\ca^{(10)}_{\mu
m_2\ldots m_6} - C^n_\mu \ca^{(10)}_{n m_2\ldots m_6})\cr
}
\eqn\efiveeight
$$
and $\hB_{m_1m_2}$ to the duals of the antisymmetric tensor fields
$$\eqalign{
\cb^{(m_1 m_2)}_{\nu\rho} =\ & {1\over 4!} \epsilon^{m_1\ldots m_6}
\ca^{(10)}_{\nu\rho m_3\ldots m_6}\cr & -[(\lambda_1 C^{m_1}_\nu C^{m_2}_\rho
+{1\over 2} \cd^{m_1}_\nu C^{m_2}_\rho -{1\over 2} \cd^{m_1}_\rho
C^{m_2}_\nu) -(m_1\leftrightarrow m_2)].\cr
}
\eqn\eneweightbb
$$

This analysis shows that the SL(2,R) invariance appears naturally in the
dimensional reduction of the dual form of the $N=1$ supergravity theory
in ten dimensions. Since the fields appearing in this form of the
supergravity theory couple naturally to the five-brane, this prompts
us
to speculate that the SL(2,R) transformation plays the same role in the
theory of five-branes as the O(6,22) transformation in the theory of
strings. The conjecture that the discrete SL(2,Z) subgroup of SL(2,R) is
an exact symmetry of string
theory\RIBANEZ\RDUALITY\RDYONO\RSCHWARZ\RDYONT\
suggests that it is an exact symmetry of the five-brane
spectrum and interactions, with the Kaluza-Klein modes and the
five-brane winding modes getting interchanged under the duality
transformation. In order to test this conjecture, however, it would be
helpful to know the full spectrum of the five-brane theory.

The similarity between the usual $R\to 1/R$ duality transformation and
the coupling constant duality transformation may be made more explicit
by expressing the
complex field $\lambda$ in terms of the variables of the dual theory.
If we define
$$
\hG_{Fmn}=G^{(10)}_{Fmn},
\eqn\efivetwentyfour
$$
then from eqs.\efivethree, \efivefive\ we get
$$
\det \hG_F=e^{-2\Phi^{(10)}}\det \hG= e^{-2\Phi}.
\eqn\efivetwentyfive
$$
This gives
$$
\lambda_2 = \sqrt{\det \hG_F}.
\eqn\efivetwentysix
$$
Combining with the first of eqs.\efiveeight\ this gives
$$
\lambda\equiv \lambda_1 +i\lambda_2 = \ca^{(10)}_{1\ldots 6} +
i\sqrt{\det \hG_F}.
\eqn\efivetwentyseven
$$
This expression is remarkably similar to the expression for the complex
field that transforms in a similar fashion under the usual target space
SL(2,Z) duality for heterotic
string compactified on a two torus:
$$
T=B^{(10)}_{12} +i\sqrt{\det\hG}.
\eqn\efivetwentyeight
$$

Our proposal fits in naturally with the observation\RDUFFLUT\ that the
roles of the $\sigma$-model loop expansion parameter and the string loop
expansion parameter get interchanged in going from the string
description of the theory to the five-brane description.
Another related observation was
made in ref.\RGAUNT, where it was found
that the magnetic monopole solutions in four-dimensional
heterotic string theory, which are crucial for the SL(2,Z)
invariance of the spectrum, may be constructed by wrapping the
five-brane soliton solutions in this theory around the six-dimensional
torus.

\chapter{Summary and Discussion}

In this paper we have shown that the low-energy effective action of
toroidally compactified heterotic string theory can be written in a
form that exhibits manifest O(6,22)$\times$SL(2,R) symmetry.
The resulting action is not manifestly general coordinate invariant, but
does possess a symmetry that reduces to the standard general coordinate
transformation laws when the auxiliary fields are eliminated by their
equations of motion.
We have also been able to get a manifestly SL(2,R) and general
coordinate invariant effective action for a restricted class of field
configurations in which all four-dimensional fields arising from the
dimensional reduction of the U(1)$^{16}$ gauge fields in ten dimensions
are set to zero. This SL(2,R) and general coordinate invariant form of
the action was shown to originate from the dimensional reduction of the
dual formulation of the $N=1$ supergravity theory in ten dimensions
without the gauge fields.

The analysis of this paper shows that the O(6,22) and SL(2,R) symmetries
appear on an equal footing from the point of view of four-dimensional
effective field theory.
Since the discrete subgroup O(6,22;Z) is known to be an exact symmetry
of (perturbative) string theory, this increases our confidence in the
conjecture that
the discrete SL(2,Z) subgroup of SL(2,R) might also be an exact symmetry
of the theory.
However, since SL(2,R) arises naturally in the dual formulation of the
ten-dimensional supergravity theory, to have the SL(2,Z) symmetry
manifest, we may need to go to a dual formulation of the theory---perhaps
the theory of five-branes.

One of the unsatisfactory features of our analysis has been that we had
to ignore the U(1)$^{16}$ gauge fields in ten dimensions to see an
SL(2,R) invariant action come out of dimensional reduction of a
ten-dimensional theory. However, since from the four-dimensional point of
view we know that a manifestly SL(2,R) invariant action of the theory
exists, one would suspect that there should be some formulation of the
$N=1$ supergravity theory in ten dimensions coupled to abelian gauge
fields, which, upon dimensional reduction, gives rise to a
manifestly SL(2,R) invariant action. Such an action would probably
provide a good starting point for the search for an alternative
formulation of the theory in which the SL(2,Z) symmetry of the spectrum
is manifest.

The analysis of sect.(2.4) already provides a clue as
to what this new formulation of the ten-dimensional supergravity theory
might be. From the analysis of sect.3,
we have seen that the action with manifest
SL(2,R) symmetry requires doubling of at least those gauge field
components that arise from the U(1)$^{16}$ gauge fields in ten
dimensions. (For the gauge fields that arise from the ten-dimensional
metric and the antisymmetric tensor fields, we can avoid the doubling,
and at the same time, maintain manifest SL(2,R) invariance, by
following the same procedure that took us from eq.\ethreetwentytwo\ to
eq.\efourfive.)
This would mean that we must have doubling of gauge fields in ten
dimensions also.
This, in turn, can be implemented by following the procedure given in
sect.(2.4).
Besides the 16 $U(1)$ gauge fields $A^I_M$, we shall now also have 16
7-form fields $\tA_{M_1\ldots M_7}^I$.
Upon dimensional reduction, the fields $A^I_M$ gives rise to scalars
$A^I_m$ and vectors $A^I_\mu$.
The fields $\tA$ give rise to vectors $\tA^I_{1\ldots 6\mu}$, which
provides the necessary doubling of the 16 U(1) gauge fields in four
dimensions.
It also gives rise to antisymmetric tensor fields $\tA_{m_1\ldots
m_5\mu\nu}$, which are dual to the scalar fields $A^I_m$ in the sense of
sect.2.4. Thus the scalar fields $A^I_m$ (which form part of the matrix
$M$) now appear in the formalism of sect.2.4. Although this destroys
manifest O(6,22) symmetry, it does not destroy manifest SL(2,R)
symmetry, since the fields $A^I_m$ (and hence also their duals)
are SL(2,R) neutral.
Finally, there are also $p$-form fields with $p\ge 3$ which appear from
the dimensional reduction of the fields $\tA$, but in four dimensions
these fields have no dynamics.

The question that one needs to address is whether there is a formulation
of $N=1$ supergravity theory coupled to abelian gauge fields that
naturally incorporates the action \etenfive\ for gauge fields in ten
dimensions.
The analysis of sect.2.5 provides a first step towards this
formulation.
If there is such a formalism, then the next question to ask would be if
this formulation of the supergravity theory arises naturally from some
other fundamental theory, possibly the five-brane, or some
generalization.

Our discussion has focussed on generic points in moduli space, where
all the gauge symmetries of the low-energy effective action are abelian.
It would be nice to extend our analysis to deal with non-abelian gauge groups.
This is a challenging problem, whose solution should be very
enlightening. Finally, let us remark that the tools that we have
introduced can be used to construct several theories that have been
sought unsuccessfully in the past. One example is a reformulation of
N=8 D=4 supergravity with the noncompact $E_{7,7}$ global symmetry
realized as a symmetry of the action. In the usual formulation this
symmetry rotates abelian gauge field strengths into their duals, just as
in the case of SL(2,R) symmetry that we have presented.
Another example is the construction of an
action for type IIB supergravity in ten dimensions.
Here the problem is the presence of a four-form potential with a
self-dual five-form field strength. As we have seen, this can also be
described by an action that sacrifices manifest covariance.

\ack
We would like to thank M. Duff, J. Harvey, C. Hull, A. Strominger and P.
Townsend for useful discussions.

\appendix

In this appendix we shall discuss the
dimensional reduction of the second term in the
action \efiveone.
We define the following four-dimensional fields in terms of the
components of the six-form potential $\ca^{(10)}$ in ten dimensions:
$$\eqalign{
\lambda_1 =\ & {1\over 6!}\epsilon^{m_1\ldots m_6} \ca^{(10)}_{m_1\ldots
m_6}\cr
\cd^{m_1}_\mu =\ & {1\over 5!} \epsilon^{m_1\ldots m_6} (\ca^{(10)}_{\mu
m_2\ldots m_6} - C^m_\mu \ca^{(10)}_{m m_2\ldots m_6})\cr
\cb^{(m_1 m_2)}_{\nu\rho} =\ & {1\over 4!} \epsilon^{m_1\ldots m_6}
\ca^{(10)}_{\nu\rho m_3\ldots m_6}\cr & -[(\lambda_1 C^{m_1}_\nu C^{m_2}_\rho
+{1\over 2} \cd^{m_1}_\nu C^{m_2}_\rho -{1\over 2} \cd^{m_1}_\rho
C^{m_2}_\nu) -(m_1\leftrightarrow m_2)]\cr
\cc^{m_1 m_2 m_3}_{\nu\rho\sigma} =\ & {1\over 3!} \epsilon^{m_1\ldots
m_6}\ca^{(10)}_{\nu\rho\sigma m_4\ldots m_6}.\cr
}
\eqn\efiveeight
$$
In terms of these fields, we define four-dimensional field strengths as
follows:
$$
F^{(\cd)m}_{\mu\nu} = \p_\mu \cd^m_\nu - \p_\nu \cd^m_\mu
\eqn\efivenine
$$
$$\eqalign{
\hK^{m_1 m_2}_{\mu\nu\rho} = & \big[\p_\mu\cb^{m_1 m_2}_{\nu\rho}
-{1\over
2}\big\{(C^{m_2}_\rho F^{(\cd)m_1}_{\mu\nu} +
\cd^{m_2}_\rho F^{(C)m_1}_{\mu\nu}) -
(m_1\leftrightarrow m_2)\big\}\big]\cr
& +{\rm cyclic ~ permutations ~ of ~} \mu, \nu, \rho\cr
}
\eqn\efiveten
$$
and
$$\eqalign{
\ck^{m_1 m_2 m_3}_{\mu\nu\rho\sigma} = & [\p_\mu \cc^{m_1 m_2
m_3}_{\nu\rho\sigma} + (-1)^P \cdot {\rm ~ cyclic ~ permutations ~ of ~}
\mu, \nu, \rho, \sigma]\cr
& - [(C^{m_3}_\sigma \hK^{m_1 m_2}_{\mu\nu\rho} +
{\rm ~ cyclic ~ permutations ~ of ~} m_1, m_2, m_3) \cr
& +(-1)^P \cdot
{\rm ~ cyclic ~ permutations ~ of ~} \mu, \nu, \rho, \sigma]\cr
& - [\{ C^{m_3}_\sigma C^{m_2}_\rho (F^{(\cd)m_1}_{\mu\nu} +\lambda_1
F^{(C)m_1}_{\mu\nu})\cr & + (-1)^P \cdot{ \rm ~ all ~ permutations ~ of ~} m_1,
m_2, m_3\}\cr
& + (-1)^P \cdot {\rm ~ inequivalent ~ permutations ~ of ~} \mu, \nu, \rho,
\sigma]\cr
& - [(C^{m_3}_\sigma C^{m_2}_\rho C^{m_1}_\nu \p_\mu \lambda_1 +
(-1)^P\cdot
{\rm ~ all ~ permutations ~ of ~} m_1, m_2, m_3)\cr
& +(-1)^P\cdot {\rm ~ cyclic ~ permutations ~ of ~} \mu, \nu,
\rho, \sigma].\cr
}
\eqn\efiveeleven
$$
In terms of these field strengths, the second term of the action
\efiveone\ may be written as
$$\eqalign{
S_2 \equiv & \int \sqrt{-\det G_F^{(10)}}e^{\Phi^{(10)}/3}(-{1\over
2\cdot
7!}) G_F^{(10)M_1 N_1}\ldots G_F^{(10)M_7 N_7} K^{(10)}_{M_1\ldots M_7}
K^{(10)}_{N_1\ldots N_7}\cr
= & -\int \sqrt{- g}\Big[ {1\over 2 (\lambda_2)^2} g^{\mu\nu}
\p_\mu\lambda_1 \p_\nu\lambda_1\cr
& +{1\over 4\lambda_2} \hG_{m_1 m_2}
g^{\mu\rho} g^{\nu\sigma} (F^{(\cd)m_1}_{\mu\nu} +\lambda_1
F^{(C)m_1}_{\mu\nu})(F^{(\cd)m_2}_{\rho\sigma}+\lambda_1
F^{(C)m_2}_{\rho\sigma})\cr
& +{1\over 2 \cdot 2! \cdot 3!}\hG_{m_1 n_1}\hG_{m_2 n_2}
g^{\mu_1\nu_1} \ldots
g^{\mu_3\nu_3}\hK^{m_1 m_2}_{\mu_1\mu_2\mu_3} \hK^{n_1
n_2}_{\nu_1\nu_2\nu_3}\cr
& + {\lambda_2\over 2 \cdot 3! \cdot 4!} \hG_{m_1 n_1}\ldots \hG_{m_3 n_3}
g^{\mu_1\nu_1}\ldots g^{\mu_4\nu_4}\ck^{m_1\ldots m_3}_{\mu_1\ldots
\mu_4} \ck^{n_1\ldots n_3}_{\nu_1\ldots \nu_4}\Big].\cr
}
\eqn\efivetwelve
$$

We would like to compare the sum of the actions
given in eqs.\efivesix\ and
\efivetwelve\ with the action given in
eq.\efourfive. In order to do this, we need to dualize the three- and
four-form field strengths $\hK^{m_1m_2}_{\mu_1\mu_2\mu_3}$ and
$\ck^{m_1\ldots m_3}_{\mu_1\ldots \mu_4}$.
We start with the four-form field strength. The equation of motion of
the field $\cc^{m_1 m_2 m_3}_{\nu_2\nu_3\nu_4}$ is given by
$$
\p_{\nu_1}[\lambda_2\sqrt{- g}\hG_{m_1 n_1}\ldots \hG_{m_3 n_3}
g^{\mu_1 \nu_1}\ldots g^{\mu_4\nu_4}\ck^{m_1\ldots m_3}_{\mu_1\ldots
\mu_4}]=0.
\eqn\efivethirteen
$$
Since $\ck^{m_1\ldots m_3}_{\mu_1\ldots \mu_4}$ is antisymmetric in
$\mu_1,\ldots \mu_4$, we may write
$$
\lambda_2 \sqrt{- g}\hG_{m_1 n_1}\ldots \hG_{m_3 n_3}
g^{\mu_1\nu_1} \ldots g^{\mu_4\nu_4} \ck^{m_1\ldots
m_3}_{\mu_1\ldots \mu_4}=\epsilon^{\nu_1\ldots \nu_4} H_{n_1 n_2 n_3}
\eqn\efivefourteen
$$
for some $H_{n_1 n_2 n_3}$.
Eq.\efivethirteen\ then gives,
$$
\p_\nu H_{n_1 n_2 n_3} =0,
\eqn\efivefifteen
$$
showing that $H_{n_1 n_2 n_3}$ is a constant.
Comparison with the original formulation of the theory shows that
$H_{mnp}$ denotes the components of the three-form field strength in the
internal directions, and hence are quantized\RROWI.
Furthermore, in the original formulation of the theory these constants
were set to zero; hence if we want to describe the same
field configurations, we must
set these constants to zero in this theory as well.
This gives
$$
\ck^{m_1 m_2 m_3}_{\mu_1\ldots \mu_4}=0.
\eqn\efivesixteen
$$

Let us now turn to the $\cb^{m_1 m_2}_{\nu_2\nu_3}$ equations of
motion.\foot{Although we are carrying out this dualization in order to
compare the dimensionally reduced action to the action \efourfive,
SL(2,R) invariance of the dimensionally reduced action can be seen
without this dualization, in the same way that in the usual scheme
O(6,22) symmetry of the action can be seen without dualizing the
$B_{\mu\nu}$ field.}
This is given by
$$
\p_{\nu_1}(\sqrt{- g}\hG_{m_1 n_1}\hG_{m_2n_2}
g^{\mu_1\nu_1}\ldots g^{\mu_3\nu_3} \hK^{m_1 m_2}_{\mu_1\mu_2\mu_3})
= 0.
\eqn\efiveseven
$$
This gives
$$
\sqrt{- g}\hG_{m_1 n_1}\hG_{m_2n_2}
g^{\mu_1\nu_1}\ldots g^{\mu_3\nu_3} \hK^{m_1 m_2}_{\mu_1\mu_2\mu_3}
=\epsilon^{\nu_1 \nu_2 \nu_3 \sigma}\p_\sigma \hB_{n_1 n_2}
\eqn\efiveeighteen
$$
for some $\hB_{n_1 n_2}$.
The Bianchi identity for the field strength $\hK^{m_1
m_2}_{\mu\nu\rho}$
$$
\epsilon^{\mu\nu\rho\sigma}\p_\sigma\hK^{m_1 m_2}_{\mu\nu\rho} =
-{3\over 2} \epsilon^{\mu\nu\rho\sigma} (F^{(C)m_1}_{\rho\sigma}
F^{(\cd)m_2}_{\mu\nu} - F^{(C)m_2}_{\rho\sigma} F^{(\cd)m_1}_{\mu\nu}),
\eqn\efivenineteen
$$
as derived from eq.\efiveten, gives rise to the following equation of
motion for the field $\hB_{m_1 m_2}$
$$
\p_\sigma (\sqrt{- g} g^{\sigma\rho}\hG^{m_1 n_1}\hG^{m_2 n_2}
\p_\rho \hB_{n_1 n_2})
={1\over 4} \epsilon^{\mu\nu\rho\sigma} (F^{(C)m_1}_{\rho\sigma}
F^{(\cd)m_2}_{\mu\nu} - F^{(C)m_2}_{\rho\sigma} F^{(\cd)m_1}_{\mu\nu}).
\eqn\efivetwenty
$$
It can be checked easily that the contributions to all the
equations of motion and Bianchi identities
derived from the term involving $\hK^{m_1 m_2}_{\mu_1\mu_2\mu_3}$ in the
action \efivetwelve\ are identical to the ones derived from the action
$$\eqalign{
-\int d^4 x\sqrt{- g}\big[&{1\over 4}\hG^{m_1 n_1} \hG^{m_2 n_2}
g^{\mu\nu} \p_\mu \hB_{m_1 m_2} \p_\nu\hB_{n_1 n_2}\cr & +{1\over
2}\hB_{m_1 m_2} F^{(C)m_1}_{\mu\nu} \tilde F^{(\cd)m_2}_{\rho\sigma}
g^{\mu\rho} g^{\nu\sigma}\big].\cr}
\eqn\efivetwentytwo
$$
Combining eqs.\efivesix, \efivetwelve, \efivesixteen\ and
\efivetwentytwo\ we get the final form of the action:
$$\eqalign{
S =  \int d^4 x \sqrt{- g} \Big[ & R -{1\over 2(\lambda_2)^2}
g^{\mu\nu}\p_\mu\lambda \p_\nu\bar\lambda \cr
& +{1\over 4} g^{\mu\nu} \p_\mu\hG_{mn} \p_\nu\hG^{mn} -{1\over 4}
g^{\mu\nu} \hG^{m_1 m_2}\hG^{n_1 n_2}\p_\mu\hB_{m_1 n_1}
\p_\nu\hB_{m_2 n_2}\cr
&-{1\over 4\lambda_2}\hG_{m_1m_2} g^{\mu\rho} g^{\nu\sigma}
\{|\lambda|^2 F^{(C)m_1}_{\mu\nu} F^{(C)m_2}_{\rho\sigma}\cr & + 2\lambda_1
F^{(C)m_1}_{\mu\nu} F^{(\cd)m_2}_{\rho\sigma} + F^{(\cd)m_1}_{\mu\nu}
F^{(\cd)m_2}_{\rho\sigma}\}\cr
& -{1\over 2} \hB_{m_1 m_2} F^{(C)m_1}_{\mu\nu} \tilde
F^{(\cd)m_2}_{\rho\sigma} g^{\mu\rho} g^{\nu\sigma}\Big].\cr
}
\eqn\efivetwentythree
$$
This action is identical to to the one given in eq.\efourfive, provided
that we identify
the vector fields $A^{(m,1)}_\mu$, $A^{(m, 2)}_\mu$
appearing in eq.\efourfive\ with $C^m_\mu$ and $\cd^m_\mu$, respectively.

\refout

\end